# SMART APPLICATION FOR AMS USING FACE RECOGNITION

MuthuKalyani.K, VeeraMuthu.A

*M-Tech Information Technology, Sathyabama University, Chennai.*
*Professor, M-Tech IT, Sathyabama University, Chennai*

## *ABSTRACT*

*Attendance Management System (AMS) can be made into smarter way by using face recognition technique, where we use a CCTV camera to be fixed at the entry point of a classroom, which automatically captures the image of the person and checks the observed image with the face database using android enhanced smart phone.*

*It is typically used for two purposes. Firstly, marking attendance for student by comparing the face images produced recently and secondly, recognition of human who are strange to the environment i.e. an unauthorized person*

*For verification of image, a newly emerging trend 3D Face Recognition is used which claims to provide more accuracy in matching the image databases and has an ability to recognize a subject at different view angles.*

## 1. INTRODUCTION

Attendance systems of old practises are not quite efficient today for keeping track on student's attendance. Due to the availability of large resources over the internet today, it is very hard to motivate the students to attend lectures without fail have become more challenging. In order to drag the attention of students and make them interactive in observing technologies we move on to latest upcoming trends on developing attendance systems. This is the strong reason for college attendance management system has to come up with an approach that ensures strong contribution of students in classrooms.

To track attendance of the students, many attendance management systems are introduced in the market. With the introduction of this variety of attendance system, skipping classes without the staff's knowledge have become difficult for the students. For few view of college attendance systems that were used earlier in the market are based on RFID systems, punch card systems, swipe card systems, biometric systems that includes fingerprint analysis, iris analysis etc. Although these systems all are lagged in their own respective so which lead to the new way practise on AMS.

In smart AMS we are going to mark the attendance of the student by capturing the image of the person for identifying correctly.

Attendance of students in the college is one of the essential day to day activities. Additional Operations within this smart system includes the software that provides

1. Marking of daily students' attendance.
2. Daily provision to check in personal attendance by employee of the college (teaching and certain non-teaching staff).
3. The software is installed to produce the attendance statistics which can be viewed by HOD, directors and staff on daily, monthly and yearly basis.

## 2. PROBLEM STATEMENT

In the standalone application [1], face was captured by the webcam cameras and the detected faces are stored in desktop webcam folder.

For matching the captured images with the database, Eigen faces methodology was used. Observance of Eigen faces method [7] was explained as,

a) Single structure of face pattern only allowed.
b) Gallery images must be of same size.
c) Requires full frontal face to be presented for each time.
d) Does not endure to the variations of brightness effect, pose and different expressions of face.
e) Effective only for low dimensional structure of face patterns.

The image which doesn't match with database is given by error message on desktop. Where the instructor has to take the attendance to those unmatched student images manually.

## 3. PROPOSED SYSTEM

The smart applications for AMS are designed as shown in the architecture diagram [Figure 1].
The CCTV camera is fixed at the entry of the class room and used to observe the face of the students. The observed image of the person is sent to the android mobile which is connected to the CCTV camera.

Android mobiles are enriched with the face recognition software technique which actually produces the possibilities of the human expression variations. The use of 3D face technology enables the work of identifying, verifying and detecting the match images in the face database.
The diagram illustrates how the phases of the system are carried out. The smart AMS consists of six modules which are explained detailed below.

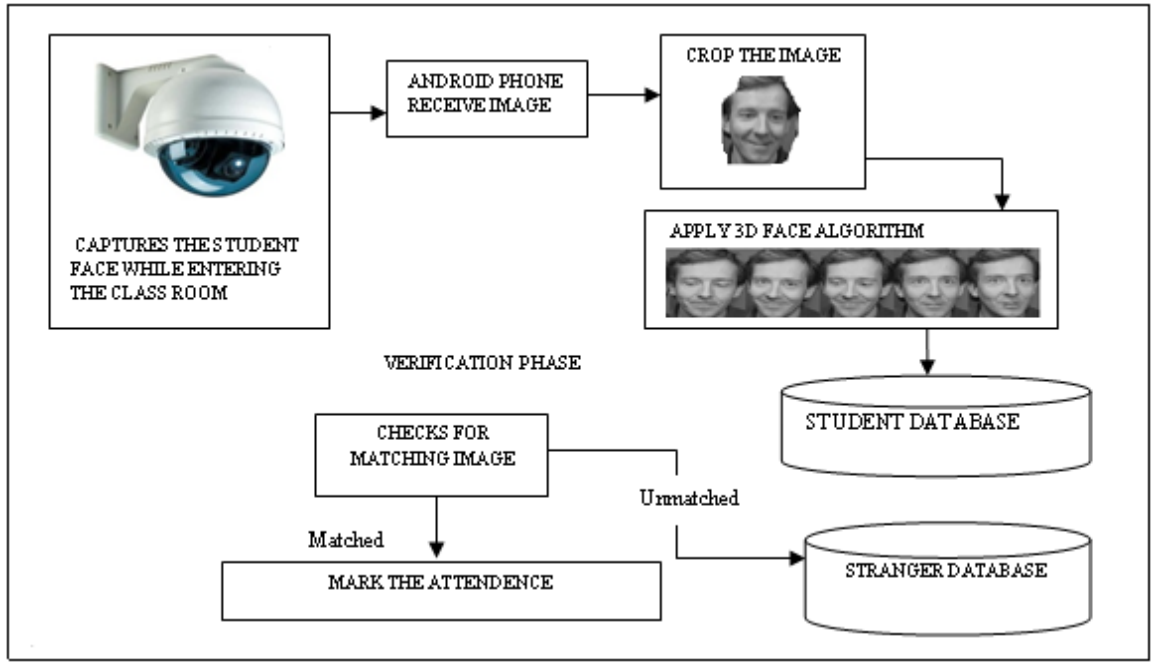

[Figure 1] Architecture diagram for smart application for AMS using Face Recognition.

### i) REGISTRATION

First phase of the system, actually deals with registering the information of the student in a particular classroom. The information includes

(1)     Name of the student
(2)     Register or Roll number
(3)     Image of the student (taken by camera).

These details are placed in the student database from which the actual comparison will be done.

### ii) IMAGE CAPTURING

The CCTV cameras are used for capturing the images [27] of the student which will be in active mode during the hours of college. The camera will be placed at the entry point of the classroom when the student enters it automatically captures the image and send to the android mobile to which it has been connected with.

The use of CCTV camera is that it is capable of capture the image of high quality and also at different angles view.

### iii) IDENTIFICATION

To identify the student image, smart phone which holds the image database of the student, checks for the match using 3D face recognition software technique.

In 3D face recognition technique [11] [15] 3D sensors are used to capture information about the outline of a face surface. The obtained information is then used to identify the distinct features on the face surface, such as the axis of the eye sockets, nose, and Chin. [Figure 2]

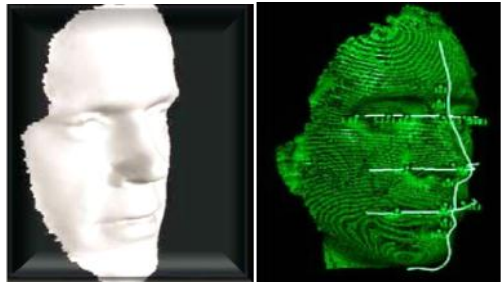

[FIGURE 2] Illustrates 3D sensors used for capturing information on the surface of a face.

Steps followed in 3D face recognition technique are:

1. Obtained image is cropped.
2. To the cropped image a 3DFace algorithm of canonical face matching is applied to get different face reactions of the particular image.

### iv) VERIFICATION

By the time of verification, dual process is done. One, the images of the students that are captured recently is compared for the match in student database [24] [28]. In two of the probabilities the images are checked. If the captured image matches with the image that has been registered before are processed for attendance management.

Second, if it is observed to be unmatched with student database then the image of the person will be consider as new and saved in the separate database called stranger database. The separation of the database will provide some information about the stranger who is new to the environment and gives the information about the person who has been entered. It not only ensure security but also make some fear to the people who needed to be entered without any authority.

### v) MARKING ATTENDANCE

The image of the student which is obtained, matched with student database and the attendance will be marked and the information is sent to the server which controls the overall database of the student. The software is installed in the smart phone that would have much additional functionality that would improve the AMS features and helps in finding the report of each student either in the percentage [Figure 3] or pie chart form as shown below in the [Figure 4]. The figure shows the overall percentage of attendance by considering the total number of working days and number of days the student attend the college.

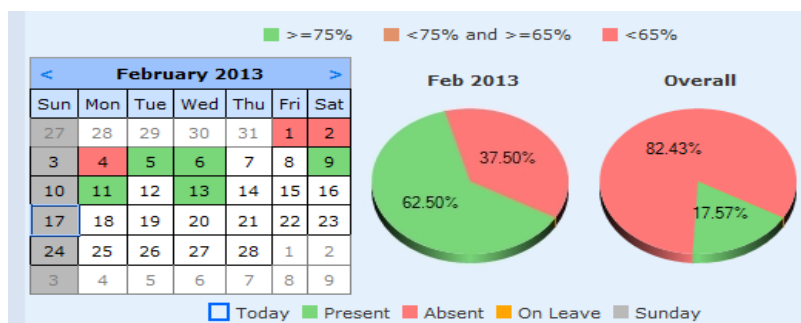

[Figure 4] Pie-chart representation of student attendance month wise.

[Figure 3] Shows the percentage of the student attended the college.

When the server receives the message of student who are absent on that particular day will send a SMS to the parent of that particular student.

### vi) SECURITY

In respect with the smart application doesn't work for AMS, it is also useful for identifying any new stranger who comes inside the classroom.

Avoids a proxy attendance of the student and ensures that the students of other class are not entering. The stranger can be the people who are not allowed to the particular environment. By which security is enhanced.

This enhanced technology development can be used in various departments of government for taking attendance for their working employees and ensure security over there.

## 4. ALGORITHM

3D Face Algorithm is a canonical face matching, which a simple and efficient surface is matching method based on high-order moments [19] [23].
The main idea is to represent the surface by its moment's $\mu_{pqr}$ up to some degree say P. which is smaller and equal to p+q+r. Compare the moments as vectors in the Euclidean space.
Given two face surfaces S1 and S2 with their corresponding canonical forms X'1 and X'2. We can define the moments-based distance between them faces as $d^{mom}(S1,S2) = \sum (\mu_{pqr}(X'_1) - \mu_{pqr}(X'_2))^2$

In practice, the vectors of moments are stored in the face database and compared to those at verification.

Example:

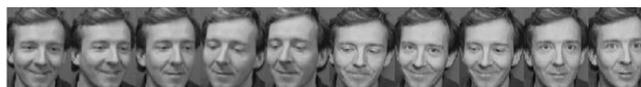

[Figure 5] Illustrates how the 3d Face Technology Applies Canonical Face matching to get different faces expressions.

17

The possibilities of face reactions are stored in the database called student database.
Main advantages of 3D face recognition are that it is not affected by any changes related to brightness of the face surface unlike other techniques practised earlier. It can also identify a face from a range of viewing away in the obtuse angled; this also includes a profile view of faces [22].

**Canonical analysis** [16] is a multivariate technique which is concerned with determining the relationships between the groups of variables in a data set. The data set is divided into two groups; let's call these groups as P and Q, based on some common characteristics defined.

The purpose of Canonical analysis is then to find the relationship between P and Q, i.e. can some form of P represent Q. It works by finding the linear combination of P variables, i.e. P1, P2 etc., and linear combination of Q variables, i.e. Q1, Q2 etc., which are most highly correlated. This combination of highly correlated variables is known as the "first canonical variants" which are usually denoted S1 and T1, with the pair of S1 and T1 being called a "canonical function". The next canonical functions, S2 and T2 are then restricted so that they are uncorrelated with S1 and T1. Everything is scaled so that the variance equals to one. The relationships between the functions are constructed which are made to agree with constraint restrictions arising from theory or to agree with common sense/experiment. These models are called as "maximum correlation models".

Mathematically, canonical analysis maximize S'P'QT subject to S'P'PS=1 and T'Q'QT=1, where P and Q are the data matrices (row for instance and column for feature).

## 5. CONCLUSION AND FUTURE ENHANCEMENT

Over last couple of years, face recognition researchers have been developing new techniques. These developments are being fuelled by advances in computer science vision techniques, computer-aided design, sensory design, and interest in the field of face recognition systems. Such advances in the various fields of interests hold the promise of reducing the error rate in face recognition systems by an order of magnitude over Face Recognition Vendor Test (FRVT).

The face recognition is grand challenges (FRGC) which are designed to achieve the performance of the goal by presenting the results to the researchers who may experiment the data produced with six-experiment challenge problem of 50,000 images. The data consists of 3D scans and high resolution still imagery taken under controlled and uncontrolled conditions.

Recognition of faces from still images is a difficult problem, because the illumination, pose and expression changes in the images create great statistical differences and the identity of the face itself becomes shadowed by these factors where the 3D face recognition has the potential to overcome feature localization, pose and illumination problems, and it can be used in conjunction with 2D systems.

Advances of the smart application plays important role in AMS and it still suffers from little deviations like the image captured using CCTV are more expensive in installing around each and every classroom of the universities and connection to android also cost less more. In case of security it does not supported for person who enters the classroom by mask the face.

1. The future work can be done on server-client application which could hold more number of images in both sides. If the server system fails the same image database can be recovered from the client side.
2. Any alternative algorithm can be used in face recognition to identify the variations of face still more clearly.

3. An effective way of analysing has to implement in identifying the people who enters the environment with masked face.
4. In all government and private offices this system can be deployed for identification, verification and attendance.